# $Cu_2ZnSnS_4$ Films using an Eco-friendly Direct Liquid Coating Approach for Solar Cell Applications

Atul Kumar[a], Anup V. Sanchela[b], C. V. Tomy[b], Ajay D Thakur[a,c]

[a]*Center for Energy and Environment, IIT Patna, Bihta 801118, INDIA*
[b]*Department of Physics, IIT Bombay, Mumbai 400076, INDIA*
[c]*Department of Physics, School of Basic Sciences, IIT Patna, Bihta 801118, INDIA*

**Abstract** $Cu_2ZnSnS_4$ (CZTS) is a promising candidate as an absorber material for thin film solar cells. The reported wet chemical synthesis approaches often have a high environmental impact due to the usage of abrasive solvents such as hydrazines, hydroxylamines, etc. We report an eco-friendly solvent based approach for making CZTS thin films with desirable absorption characteristics for solar cell applications.

## 1. Introduction

With increasing energy demand, cost of energy becomes an important factor [1]. The current need of energy at low cost is one of the many driving forces for the renewable energy industry. $CuIn_xGa_{1-x}S_2$ (CIGS) absorber layer based photovoltaic (PV) technology with an efficiency of 21.7% [2] appeared to be the best low cost replacement for expensive Si PV technology (efficiency 25.6%) [3]. However critical issues including In price fluctuation and the toxicity associated with Ga processing severely hampered its prospects. In view of this, $Cu_2ZnSnS_4$ (CZTS) presents itself as a viable alternative [4] to CIGS based PV technology. CZTS has an expanded chalcopyrite structure where the trivalent In/Ga is replaced by bivalent Zn and tetravalent Sn. The primary methods used for fabricating CZTS thin films include co-evaporation [5], PLD [6], sputtering [7], CVD [8], electrodeposition [9], SILAR [10], photochemical deposition [11], solvothermal [12], ball milling/solid state route [13], microwave [14], inert-atmosphere based chemical route [15], sol-gel [16], spray pyrolysis [17], nanoparticle [18] and nanoink [19] based techniques. The environmental impact of the above processes in light of energy input and processing techniques along with infrastructural requirements and scalability issues are detrimental to the success of CZTS based PV technologies. For example, sputtering, PLD and co-evaporation suffer from issues like high energy input, non-scalability and intensive infrastructural costs, phase formation is a critical issue with electrochemical synthesis and co-evaporation. The reported wet chemical approaches for making CZTS thin films suitable for solar cell have a high environmental impact due to the usage of abrasive solvents such as hydrazines, hydroxylamines, etc. Scalable techniques with low energy input and nontoxic precursor/solvent with versatile deposition of phase selective CZTS films is an indispensable requirement. Solution processed CZTS film have advantages of low cost, high throughput, less energy intensive over non vacuum techniques. Solution/liquid processed techniques can be broadly characterized into two categories: (i) nanoink based approach where nanoparticles of kestrite CZTS are dispersed in a volatile solvent, and (ii) direct liquid coating (DLC) based approach where molecular precursors in solvent form a correct phase upon annealing. Good account of work is reported in the reviews by Romanyuk *et al* [20] and Abbermann [21]. The highest efficiency solar cells based on CZTS are solution processed but they utilize toxic solvents like hydrazine (record efficiency of 12.6 [1]) and more recently hexylamine [22]. Other solvents such as DMSO [23], methoxyethanol [24], pyridine [25], ethylcellulose [26] have been reported which too have a considerable degree of toxicity. De-ionised water-ethanol mixture [27] and ethanol-methanolamine mixture [28] are reported with relatively less toxicity for solution processed CZTS, but stability of solution is an issue. In this respect a safer and stable solvent becomes an important step to be addressed for direct solution processed CZTS film. Here we report a comparatively eco-friendly solvent system, viz., isopropanol-polyethylene glycol for synthesis of solution processed CZTS using DLC technique.

## 2. Experiment details

Thin films were deposited by DLC approach using isopropanol and polyethylene glycol as solvents respectively. For this, precursors $CuCl_2$, $ZnCl_2$, $SnCl_4$, and a sulphur source (thioacetamide) are used in ratio 1.5:1.02:1.8:5.5. Copper, Zinc and Tin salts are dissolved in isopropanol to obtain a greenish solution. Thioacetamide is dissolved in polyethylene glycol and stirred to form a yellowish solution. A small amount of methanolamine is added to this solution. Under stirring the isopropanol solution is added to the latter solution to form a molecular precursor solution (labeled OT) and stored for drop casting/spin coating deposition of CZTS films. For the other synthesis, $CuCl_2$, $ZnCl_2$, $SnCl_2$ in ratio 2:1:1 was dissolved in isopropanol. To this, 3mM thioacetamide is added with methanolamine resulting in another molecular precursor solution (labeled ST). After coating of films using the two precursor solutions, annealing is done at $220^0C$ in open air for 30 minutes. Data for films labeled OT24 and ST21 grown using precursors OT and ST, respectively are reported here. The crystallographic study is done using Rigaku TTRX-IV XRD employing Cu K. The Raman spectroscopy is done using the Jobin-Yvon Raman setup using 632.8 nm He:Ne laser in the micro-Raman configuration. EDAX for Elemental composition and SEM for surface morphology are done by Hitachi 480 with accelerating voltages of 15kV and 10kV, respectively.

## 3. Result and discussion

Figure 1 (a) shows the x-ray diffraction (XRD) data for the film ST21. Prominent peaks at $28.28^0$, $47.38^0$, $56.34^0$ are indexed to 112, 220, 312 peaks of the kesterite CZTS phase (ICDD 01-075-4122). As the binary compound $Cu_2S$ (ICDD 01-073-6078) has XRD peaks located at overlapping locations, XRD alone is not a conclusive proof for the formation of CZTS. Furthermore, there is a large propensity for the formation of $Cu_2S$ as a major phase during synthesis of CZTS. Thus Raman measurement is done to declassify any binary present in sample and confirm its phase purity. The Raman data for ST21 is shown in Fig. 1 (b) along with the multi-peak fits. Raman peaks at 272 $cm^{-1}$, 333 $cm^{-1}$ corresponds to CZTS [29-31]. The minor peak at 424 $cm^{-1}$ corresponds to $Cu_2S$ indicating the presence of a secondary $Cu_2S$ phase [32]. This peak can be eliminated by further optimization of synthesis process, like annealing and sulfurisation. The EDAX data show the elemental composition as 2.4:1:1.03:4.07 (see Fig. 1(c)). The Cu/Zn+Sn ratio is 1.19 which suggests it to be a Cu rich, Zn poor film. The final film is continuous as shown in the SEM image in Fig. 1 (d). Figure 2 (a) shows the x-ray diffraction (XRD) data for the film OT24. The major peaks at $28.34^0$, $47.36^0$, $56.02^0$ $34^0$ are indexed to 112, 220, 312 peaks of the kesterite CZTS phase (ICDD 01-075-4122). A minor peak at $14.32^0$ corresponding to ternary phase $Cu_3SnS_4$ is seen. This is seen to get suppressed upon increasing the annealing temperature. Figure 2 (b) shows the Raman data for OT24. The multi-peak fit shows de-convoluted Raman peaks at 266 $cm^{-1}$, 295 $cm^{-1}$, 332.7 $cm^{-1}$, 365 $cm^{-1}$ and 435$cm^{-1}$. The peaks at 266 $cm^{-1}$, 332.7 $cm^{-1}$, 365 $cm^{-1}$ corresponds to CZTS [29-31]. Peak at 295 $cm^{-1}$ corresponds to the ternary phase $Cu_3SnS_4$ [33]. The EDAX data shows elemental a composition of 2.7:4.21:1:4.24 (see Fig. 2(c)). This corresponds to a Cu poor sample with a Cu/Zn+Sn ratio of 0.52. The drop casted film is continuous and well connected as shown in SEM image in Fig. 2 (d). The current solvent system showed a degree of flexibility in optimization the composition ratio. The films are open air annealed at $220^0C$ and are not further sulfurised. The minor binary and ternary impurity phases could be eliminated using suitable sulfurization and work is under progress in this direction.

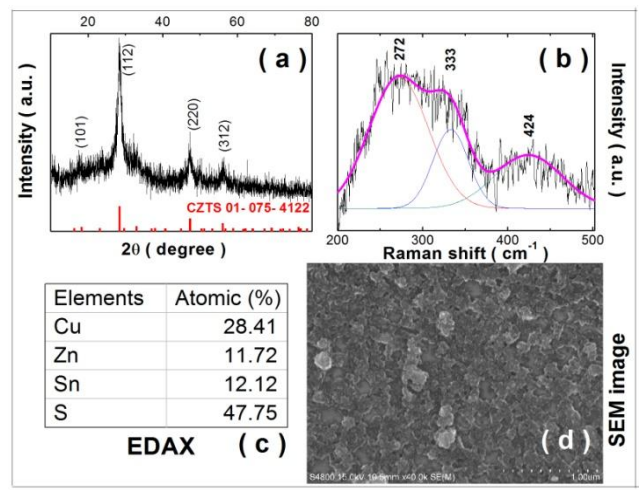

Fig. 1. The XRD, Raman, EDAX and SEM graph of sample ST21.

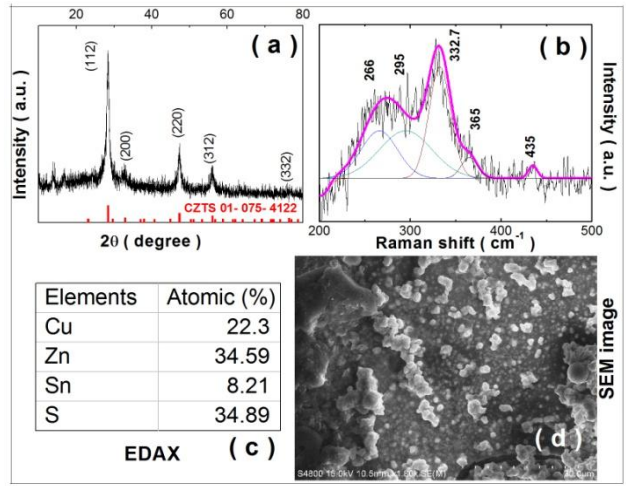

Fig. 2. The XRD, Raman, EDAX and SEM graph of sample OT24.

## 4. Conclusion

The presented synthesis technique shows that direct coating with non toxic solvent mixture can be used to deposit CZTS films. The film processing involves a low temperature non-vacuum method which is easily scalable. The benign solvent can increase the acceptability of the CZTS thin film fabrication route. Further work is in progress for using these thin films in solar cell applications.